\documentclass[twocolumn,preprintnumbers,amsmath,amssymb,showpacs]{revtex4}

\usepackage{graphicx}
\usepackage{dcolumn}
\usepackage{bm}
\usepackage{theorem}
\usepackage{color}
\usepackage{multirow}

{\theorembodyfont{\normalfont}
\theoremheaderfont{\normalfont\it}
\newtheorem{thm}{ Theorem}
\newtheorem{dfn}[thm]{ Definition}
\newtheorem{lmm}[thm]{ Lemma}

}

{\theorembodyfont{\normalfont}
\theoremheaderfont{\normalfont\it}
\newtheorem{prf}{ Proof:}}

{\theorembodyfont{\normalfont}
\theoremheaderfont{\normalfont\it}
}

{\theorembodyfont{\normalfont}
\theoremheaderfont{\normalfont\it}
}

{\theorembodyfont{\normalfont}
\theoremheaderfont{\normalfont\it}
}

{\theorembodyfont{\normalfont}
\theoremheaderfont{\normalfont\it}
\newtheorem{rmk}{ Remark.}}

\newcommand{\bra}[1]{\mbox{$\left\langle#1\right|$}}
\newcommand{\ket}[1]{\mbox{$\left|#1\right\rangle$}}

\newcommand{\proj}[1]{\mbox{$\ket{#1}\!\bra{#1}$}}

\newcommand{\alg}[1]{\begin{align}#1\end{align}}
\newcommand{\rarrow}{\rightarrow}

\newcommand{\ca}[1]{{\mathcal #1}}
\newcommand{\Tr}[2]{{\rm Tr}_{#1}\left[#2\right]}
\newcommand{\bthm}[1]{\begin{thm}\label{thm:#1}}
\newcommand{\ethm}{\end{thm}}
\newcommand{\rthm}[1]{Theorem \ref{thm:#1}}
\newcommand{\blmm}[1]{\begin{lmm}\label{lmm:#1}}
\newcommand{\elmm}{\end{lmm}}
\newcommand{\bdfn}[1]{\begin{dfn}\label{dfn:#1}}
\newcommand{\edfn}{\end{dfn}}
\newcommand{\bprf}{\begin{prf}}
\newcommand{\eprf}{\end{prf}}
\newcommand{\brmk}{\begin{rmk}}
\newcommand{\ermk}{\end{rmk}}

\newcommand{\rineq}[1]{Inequality (\ref{eq:#1})}
\newcommand{\QED}{\hfill$\blacksquare$}
\newcommand{\lsec}[1]{\label{sec:#1}}
\newcommand{\rsec}[1]{\ref{sec:#1}}

\newcommand{\nn}{\nonumber}
\newcommand{\beq}{\begin{eqnarray}}
\newcommand{\eeq}{\end{eqnarray}}

\begin{document}


\title{Symmetrizing Cost of Quantum States}

\author{Eyuri Wakakuwa}
\email{wakakuwa@quest.is.uec.ac.jp}
\affiliation{Department of Communication Engineering and Informatics, Graduate School of Informatics and Engineering, The University of Electro-Communications, Japan}%

\date{\today}

\begin{abstract}
We introduce and analyze a task that we call {\it symmetrization}, in which a state of a quantum system, associated with a symmetry group, is transformed by a random unitary operation to a symmetric state. Each element of the unitary ensemble is required to be symmetry preserving, in the sense that it keeps the set of symmetric states invariant. We consider an asymptotic limit of infinitely many copies and vanishingly small error, and analyze the {\it symmetrizing cost}, that is, the minimum cost of randomness per copy required for symmetrization. We prove that the symmetrizing cost of an arbitrary quantum state is equal to the relative entropy of frameness, thereby providing it with a direct operational meaning.
\end{abstract}

                       
                                                      
\maketitle

\section{Introduction}

The concepts of symmetry and asymmetry, which have played significant roles in the development of modern physics\cite{gross1996role}, have recently been attracting much attention in the field of quantum information theory. For one reason, this is because our ability to perform quantum information processing tasks is in general limited by symmetry of physical systems (see e.g. \cite{bartlett07}). For the other reason, it is because the concepts and techniques developed in quantum information theory are useful to analyze the notions of symmetry and asymmetry from an operational viewpoint\cite{piani16,gilad08,marvian2013theory,marvian2014asymmetry,marvian14,marvian2012information,ahmadi2013wigner}.

\begin{figure}[t]
\begin{center}
\includegraphics[bb={0 10 490 316}, scale=0.45]{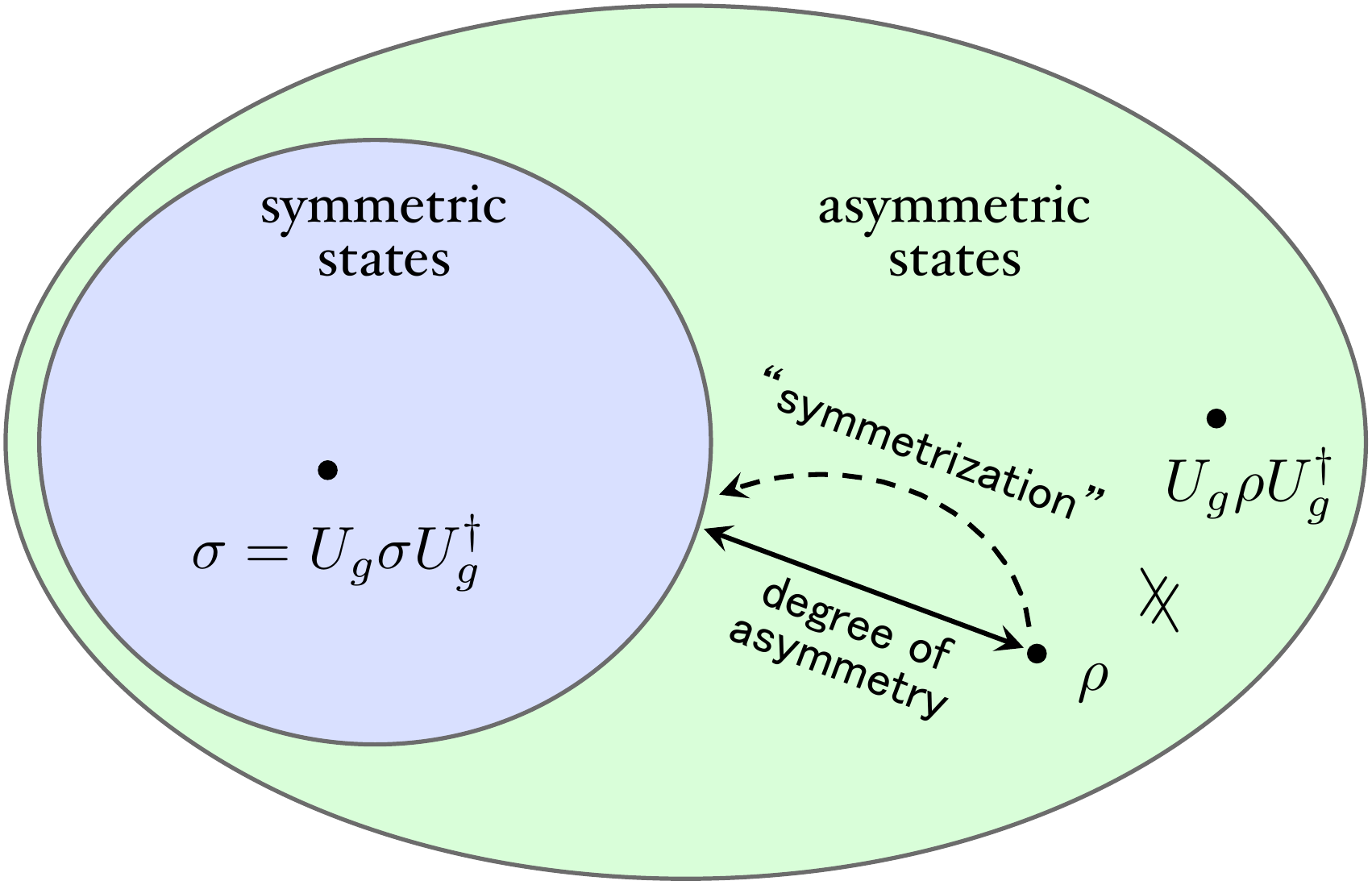}
\end{center}
\caption{A schematic diagram of the task of symmetrization is depicted. The symmetrizing cost, defined as the minimum cost of randomness required for symmetrization, quantifies the degree of asymmetry of quantum states operationally. Note that a state $\sigma$ is defined to be symmetric if it satisfies $U_g\sigma U_g^\dagger=\sigma$ for all $g\in G$, where $G$ is a symmetry group with a unitary representation $\{U_g\}_{g\in G}$.
}
\label{fig:A}
\end{figure}

A central problem lying in these studies is how to quantify the degree of asymmetry of quantum states\cite{gour07,piani16,gilad08,vaccaro2008tradeoff,skotiniotis12,toloui2011constructing,marvian2013theory,marvian14,marvian2014asymmetry}. Mathematically, this problem reduces to a search for functions that are monotonically nonincreasing under operations restricted by symmetry\cite{gour07}. The {\it relative entropy of frameness} (REF)\cite{gilad08} is such a function, which is applicable for general symmetry groups. The REF is straightforwardly computable, and has partial operational interpretations in the context of thermodynamic work extraction\cite{vaccaro2008tradeoff}, reference frame sharing\cite{gilad08,skotiniotis12} and quantum metrology\cite{hall12}. However, a direct operational meaning of the REF is yet unknown.

In this paper, we answer the above problem by introducing a task that we call {\it symmetrization}, in which an asymmetric state is transformed by a random unitary operation to a symmetric state (FIG.\ref{fig:A}). Each element of the unitary ensemble is required to be {\it symmetry preserving}, i.e., it maps any symmetric state to another, or equivalently, it keeps the set of symmetric states invariant. We consider an asymptotic limit of infinitely many copies and vanishingly small error, and analyze the minimum cost of randomness per copy required for symmetrization. The minimum cost is referred to as the {\it symmetrizing cost}. We prove that the symmetrizing cost is equal to the REF, thereby providing a direct operational meaning to the REF. Our result, as a special case, reduces to that of \cite{singh15}, which analyzed the minimum cost of randomness required for destroying quantum coherence.

This paper is organized as follows. Section \ref{sec:symmetry} reviews mathematical treatments of symmetry and asymmetry of quantum states. Section \rsec{symmetrizingcost} introduces the formal definitions of symmetrization and the symmetrizing cost, and describes the main result. Conclusions are given in Section \ref{sec:conclusion}. A proof of the main result is provided in Appendix \ref{app:proof}. Throughout this paper, we denote by ${\mathcal S}({\mathcal H})$ the set of normalized state on a Hilbert space $\ca H$, and by $A^n$ a system composed of $n$ identical systems of $A$. $\log{x}$ represents the base $2$ logarithm of $x$.

\section{Symmetry of Quantum States}\label{sec:symmetry}

In this section, we review mathematical definitions of symmetry and asymmetry of quantum states. We also introduce a function known as the {\it relative entropy of frameness}.

\subsection{Single System}

Let us consider a quantum system $A$ described by a Hilbert space ${\ca H}^A$ with dimension $d\:(<\infty)$, and consider a group $G$ of physical transformations on $A$. Following the previous literatures\cite{bartlett07,piani16,gilad08,marvian2013theory,marvian2014asymmetry,marvian14,marvian2012information,ahmadi2013wigner,gour07,vaccaro2008tradeoff,skotiniotis12,toloui2011constructing,hall12}, we assume that (i) $G$ has a unitary representation $\{U_g\}_{g\in G}$ on ${\ca H}^A$, and that (ii) $G$ is a compact Lie group or a finite group. 

A state $\sigma$ is defined to be {\it symmetric with respect to $G$} if $U_g\sigma U_g^\dagger=\sigma$ for all $g\in G$. Conversely, a state $\rho$ is said to be {\it asymmetric with respect to $G$} if $U_g\rho U_g^\dagger\neq\rho$ for at least one element $g\in G$. We denote the set of symmetric states by ${\ca S}_{\rm sym}(G)$. We define a unitary $V$ acting on ${\ca H}^A$ to be {\it symmetry preserving with respect to $G$} if it keeps the set of symmetric states invariant, or equivalently, if
\begin{align}
V\sigma V^\dagger\in{\ca S}_{\rm sym}(G),\;\forall \sigma\in {\ca S}_{\rm sym}(G).\nn
\end{align} 
In the rest of this paper, we fix an arbitrary group $G$ satisfying the above conditions (i) and (ii), and omit the phrase ``with respect to $G$'' when there is no fear of confusion. The following lemma immediately follows:
\blmm{symconv}
For any $\sigma\in{\ca S}_{\rm sym}(G)$ and $\lambda\in(0,1]$, we have $(1-\lambda)\sigma+\lambda\rho\in{\ca S}_{\rm sym}(G)$ if and only if $\rho\in{\ca S}_{\rm sym}(G)$. Hence ${\ca S}_{\rm sym}(G)$ is a convex set.
\elmm

The {\it twirling operation} $\ca{T}_G$ on system $A$ with respect to $G$ is defined for a compact Lie group by
\alg{
\ca{T}_G(\tau)=\int_GdgU_g\tau U_g^\dagger,\;\forall\tau\in{{\ca S}({\ca H}^A)},\label{eq:defgtwirl2}
}
with $dg$ being the group invariant (Haar) measure on $G$, and for a finite group by
\alg{
\ca{T}_G(\tau)=\frac{1}{|G|}\sum_{g\in G}U_g\tau U_g^\dagger,\;\forall\tau\in{{\ca S}({\ca H}^A)},\label{eq:defgtwirl}
}
with $|G|$ denoting the order of $G$. In both cases, we have $\ca{T}_G(\sigma)=\sigma$ for any symmetric state $\sigma$ by definition. Any state is mapped to a symmetric state by $\ca{T}_G$, i.e., $\ca{T}_G(\rho)\in{\ca S}_{\rm sym}(G)$ for all $\rho\in{\ca S}({\ca H})$, since we have
\alg{
U_g\ca{T}_G(\rho)U_g^\dagger=\ca{T}_G(\rho),\;\forall\rho\in{{\ca S}({\ca H}^A)},\forall g\in G.\nn
}

\subsection{Composite System}\label{sec:symconp}

Let us extend the above definition of symmetry to that on a composite system (see also Appendix \ref{app:remark}). Consider a system composed of $n$ duplicates of $A$, which we denote by $A^n=A_1\cdots A_n$. It would be natural to define that a state $\sigma$ on $A^n$ is symmetric if it is invariant under the action of symmetry group $G$ on any of $A_1,\cdots, A_n$. Hence we consider a symmetry group $G^{\times n}=G\times\cdots\times G$, where $\times$ represents the direct product of groups. Denoting $(g_1,\cdots,g_n) \in G^{\times n}$ by ${\vec g}$ and $U_{g_1}\otimes\cdots\otimes U_{g_n}$ by  $U_{\vec g}$, we define a state  $\sigma\in{\mathcal S}(({\mathcal H}^A)^{\otimes n})$ to be {\it symmetric} if it satisfies
\alg{
U_{\vec g}\sigma U_{\vec g}^{\dagger}=\sigma,\quad\forall{\vec g}\in G^{\times n}.\label{eq:ugsug}
}
The set of symmetric states on $A^n$ is denoted by ${\ca S}_{\rm sym}(G^{\times n})$.

Similarly to the case of a single system, we define a unitary $V$ acting on $({\ca H}^A)^{\otimes n}$ to be {\it symmetry preserving} if it satisfies
\begin{align}
V\sigma V^\dagger\in{\ca S}_{\rm sym}(G^{\times n}),\;\forall \sigma\in {\ca S}_{\rm sym}(G^{\times n}).\nn
\end{align}
By definition, $U_{\vec g}$ is symmetry preserving for all ${\vec g}\in G^{\times n}$. The twirling operation $\ca{T}_{G^{\times n}}$ is expressed by 
\alg{
\ca{T}_{G^{\times n}}(\tau)=\int_Gd{\vec g}\:U_{\vec g}\tau U_{\vec g}^\dagger,\label{eq:juji}
}
where we denoted $dg_1\cdots dg_n$ by $d{\vec g}$, with $dg_i$ being the $G$-invariant (Haar) measure for each $i$. It is straightforward to verify that
\alg{
\ca{T}_{G^{\times n}}=\ca{T}_G^{\otimes n}.\label{eq:jujiiii}
}
Again $\ca{T}_{G^{\times n}}(\rho)$ is a symmetric state for any state $\rho$, and $\ca{T}_{G^{\times n}}(\sigma)=\sigma$ for any symmetric state $\sigma$.

\subsection{Relative Entropy of Frameness}

To quantify the degree of asymmetry of quantum states, Gour et al.\cite{gilad08} introduced a function called the {\it relative entropy of frameness} (REF), defined by
\begin{align}
D_G(\rho):=\min_{\sigma\in{\ca S}_{\rm sym}(G)}D(\rho\|\sigma)\label{eq:defrg}
\end{align}
with the quantum relative entropy $D(\rho\|\sigma):={\rm Tr}[\rho\log{\rho}]-{\rm Tr}[\rho\log{\sigma}]$. They proved that the REF have a simple expression in terms of the von Neumann entropy, namely, that
\begin{align}
D_G(\rho)=D(\rho\|\ca{T}_G(\rho))=S(\ca{T}_G(\rho))-S(\rho)\label{eq:rgeqsgmns}
\end{align}
for all $\rho\in{\ca S}({\ca H}^A)$ and $S(\rho):=-{\rm Tr}[\rho\log{\rho}]$.

The REF has partial operational meanings: It provides upper bounds on the amount of thermodynamical work, extractable from a state under the restriction by superselection rules\cite{vaccaro2008tradeoff}, on the performance of an asymmetric state to act as an indicator of a reference frame\cite{gilad08,skotiniotis12}, and on the estimation error of phase shifts in quantum metrology\cite{hall12}. However, a direct operational meaning of the REF is still unknown, which is the subject solved in this paper.

\begin{figure}[t]
\begin{center}
\includegraphics[bb={0 0 296 210}, scale=0.55]{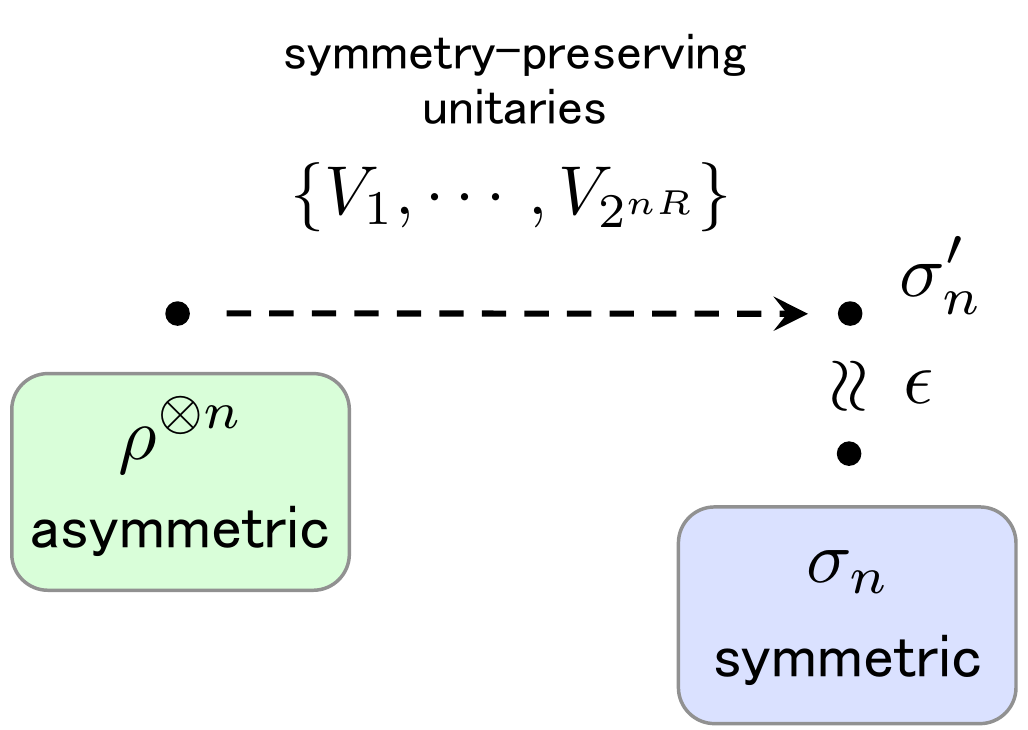}
\end{center}
\caption{A schematic diagram of the task of symmetrization is depicted. $n$ copies of an asymmetric state $\rho$ on system $A$ is transformed by a random application of symmetry preserving unitaries $\{V_1,\cdots,V_{2^{nR}}\}$ on $A^n$ with the uniform distribution. We require that the state after the transformation is a symmetric state $\sigma_n$ up to a small error $\epsilon$. The symmetrizing cost of $\rho$ is defined as the minimum rate $R$ such that $\epsilon\rightarrow0$ can be accomplished in the limit of $n\rightarrow\infty$, by properly choosing $\{V_1,\cdots,V_{2^{nR}}\}$ and $\sigma_n$ for each $n$.}
\label{fig:B}
\end{figure}

\section{Symmetrizing Cost}\label{sec:symmetrizingcost}

Consider an asymmetric state $\rho$ on system $A$. Suppose one wants to transform $\rho$ to a symmetric state, i.e. to {\it symmetrize} $\rho$, by randomly applying certain symmetry preserving unitaries. This is always possible, since Equality (\ref{eq:defgtwirl2}) or (\ref{eq:defgtwirl}) holds and $U_g$ is symmetry preserving for any $g\in G$. But what if at the same time one also wants to minimize the number of unitaries, or equivalently, to minimize the cost of randomness, for symmetrizing $\rho$?

Intuitively, it would be natural to expect that if $\rho$ is close to (but not equal to) a symmetric state, the minimum cost of randomness required for symmetrizing it is small. However, this is not the case: Consider a symmetric state $\sigma$, an asymmetric state $\rho$ and define an asymmetric state
\alg{
\varrho(\lambda):=(1-\lambda)\sigma+\lambda\rho,\quad\lambda\in(0,1].\nn
}
Let $\{V_k\}_{k=1}^K$ be a set of symmetry preserving unitaries such that we have
\alg{
{\ca R}(\varrho(\lambda))\in{\ca S}_{\rm sym}(G)\nn
}
for
\alg{
{\ca R}:\tau\rightarrow\frac{1}{K}\sum_{k=1}^KV_k\tau V_k^\dagger.\nn
}
By definition, we have $V_k\tau V_k^\dagger\in{\ca S}_{\rm sym}(G)$ for each $k$. Due to the convexity of ${\ca S}_{\rm sym}(G)$, this implies ${\ca R}(\sigma)\in{\ca S}_{\rm sym}(G)$. Applying Lemma \ref{lmm:symconv} and noting that ${\ca R}(\varrho(\lambda))=(1-\lambda){\ca R}(\sigma)+\lambda{\ca R}(\rho)$, we obtain ${\ca R}(\rho)\in{\ca S}_{\rm sym}(G)$. Hence the cost of randomness required for symmetrizing $\varrho(\lambda)$ is as large as one required for symmetrizing $\rho$, despite the fact that asymmetry of $\varrho(\lambda)$ can be much smaller than that of $\rho$ when $\lambda$ is small.

To circumvent such a mismatch, we consider an asymptotic limit of infinitely many copies and vanishingly small error. That is, we consider a task in which $n$ copies of $\rho$ on system $A$ is transformed to a symmetric state on $A^n$ by randomly applying certain symmetry preserving unitaries over $A^n$ (FIG.\ref{fig:B}). We do not require that the state after the operation is {\it exactly} a symmetric state for finite $n$. Instead, we require that the state is equal to a symmetric state {\it within a small error} for large $n$, and that the error vanishes in the limit of $n\rightarrow\infty$. The {\it symmetrizing cost} is defined as the minimum cost of randomness per copy required for accomplishing this task. A rigorous definition of the symmetrizing cost is given as follows:

\bdfn{symcost}
A rate $R$ is said to be achievable in symmetrizing a state $\rho\in{\mathcal S}({\mathcal H}^A)$ with respect to $G$ if, for any $\epsilon>0$ and sufficiently large $n$, there exist a symmetric state $\sigma_n$ on $A^n$ and a set of symmetry preserving unitaries $\{V_k\}_{k=1}^{2^{nR}}$ on $A^n$ such that
\alg{
\left\|{\ca V}_n(\rho^{\otimes n})-\sigma_n\right\|_1\leq\epsilon\nn
}
for
\alg{
\ca{V}_n:\tau\rarrow\frac{1}{2^{nR}}\sum_{k=1}^{2^{nR}}V_{k}\tau V_{k}^\dagger.\nn
}
The symmetrizing cost of a state $\rho\in{\mathcal S}({\mathcal H}^A)$ with respect to $G$ is defined as \alg{
&C_G(\rho):=\nn\\
&\inf\{R\:|\:R\text{ is achievable in symmetrizing }\rho\text{ w.r.t. }G\}.\nn
}
\edfn

The main result of this paper, represented by the following theorem, is that the symmetrizing cost of a state is equal to the relative entropy of frameness. A proof will be given in Appendix \ref{app:proof}.
\bthm{symmetrizingcost}
For an arbitrary $\rho\in{\ca S}({\ca H}^A)$, it holds that
\alg{
C_G(\rho)=D_G(\rho).\nn
}
\ethm

\section{Conclusion} \label{sec:conclusion}

In this paper, we introduced the task of symmetrization, and analyzed the minimum cost of randomness required for symmetrizing a quantum state. We particularly considered an asymptotic limit of infinitely many copies and vanishingly small error. We proved that the minimum cost of randomness is asymptotically equal to the relative entropy of frameness, thereby providing it with a direct operational meaning. To find applications of our results, e.g. to quantum information processing tasks, is left as a future work. 

A general idea behind the concept of the symmetrizing cost is that we can quantify a certain property of quantum states in terms of the cost of randomness required for erasing it. This idea was first introduced in \cite{berry05} for quantifying multipartite quantum correlations, inspired by Landauer's principle\cite{landauer}. The recent studies apply the same idea to quantify non-Markovianity of tripartite quantum states\cite{wakakuwa2016markovianizing,waka15_rec,deconstruction} and to quantify quantum coherence\cite{singh15}. We conclude this paper by pointing out that it would be fruitful to further explore applicability of this idea to other properties of quantum states, such as athermality in quantum thermodynamics\cite{gour2015resource}, non-Gaussianity in quantum optics\cite{browne2003driving}, and contextuality\cite{grudka2014quantifying} and steerability\cite{gallego2015resource} in the foundation of quantum mechanics.

\begin{acknowledgments}
\noindent The authors thank A. Winter for helpful discussions, and H. Tajima, D. Asakura for useful comments on earlier drafts of this paper.
\end{acknowledgments}

\hfill


\appendix

\section{Mathematical Preliminaries}

In this appendix, we review technical tools that will be used to prove Theorem \ref{thm:symmetrizingcost} in the following appendix. For the references, see e.g. \cite{nielsentext,wildetext,cover05}.

\subsection{Trace Distance and The Gentle Measurement Lemma}\label{app:TDGML}

The trace norm of a linear operator $A$ on a finite dimensional Hilbert space is defined by $\|A\|_1:={\rm Tr}[\sqrt{A^\dagger A}]$. 
The trace distance between two subnormalized states $\rho$ and $\sigma$ is defined by
\alg{
\|\rho-\sigma\|_1={\rm Tr}[\sqrt{(\rho-\sigma)^2}].\nn
}
The trace distance is monotonically nonincreasing under quantum operations, i.e., it satisfies
\begin{eqnarray}
\left\|\rho-\sigma\right\|_1\geq\left\|{\mathcal E}(\rho)-{\mathcal E}(\sigma)\right\|_1\nn
\end{eqnarray}
for any linear CPTP map ${\mathcal E}$. For three subnormalized states $\rho,\sigma$ and $\tau$, we have
\begin{eqnarray}
\left\|\rho-\tau\right\|_1\leq\left\|\rho-\sigma\right\|_1+\left\|\sigma-\tau\right\|_1,\nn
\end{eqnarray}
which is called the {\it triangle inequality}.   


The gentle measurement lemma (Lemma 2 in \cite{berry05}) states that for any subnormalized density operator $\rho$ and linear operator $X$ on a finite dimensional Hilbert space $\ca H$, and for any $\epsilon\geq0$ such that $0\leq X\leq I$ and ${\rm Tr}[\rho X]\geq1-\epsilon$, we have
\begin{eqnarray}
\|\rho-\sqrt{X}\rho\sqrt{X}\|_1\leq2\sqrt{2\epsilon}.\nn
\end{eqnarray}
The following lemma is obtained as a corollary:
\blmm{ggentle}
Let $\rho$ and $\sigma$ be arbitrary subnormalized density operators on $\ca H$, and let $\Pi$ be a projection onto a subspace of $\ca H$ such that
\alg{
&\Tr{}{\Pi\rho\Pi}\geq1-\epsilon_1,\nn\\
&\left\|\Pi\rho\Pi-\Pi\sigma\Pi\right\|_1\leq\epsilon_2\label{eq:pirhopis}
}
for $\epsilon_1,\epsilon_2\in[0,1]$. Then we have
\alg{
\left\|\rho-\sigma\right\|_1\leq6\sqrt{2(\epsilon_1+\epsilon_2)}.\nn
}
\elmm

\bprf
By the triangle inequality for the trace norm, we have
\alg{
\Tr{}{\Pi\sigma\Pi}&=\left\|\Pi\sigma\Pi\right\|_1\nn\\
&\geq\left\|\Pi\rho\Pi\right\|_1-\left\|\Pi\rho\Pi-\Pi\sigma\Pi\right\|_1\nn\\
&=\Tr{}{\Pi\rho\Pi}-\left\|\Pi\rho\Pi-\Pi\sigma\Pi\right\|_1\nn\\
&\geq1-\epsilon_1-\epsilon_2.\nn
}
The gentle measurement lemma yields
\alg{
&\left\|\rho-\Pi\rho\Pi\right\|_1\leq2\sqrt{2\epsilon_1},\nn\\
&\left\|\sigma-\Pi\sigma\Pi\right\|_1\leq2\sqrt{2(\epsilon_1+\epsilon_2)}.\nn 
}
Applying the triangle inequality to (\ref{eq:pirhopis}) and the above two inequalities, we obtain that
\alg{
\left\|\rho-\sigma\right\|_1\leq\epsilon_2+2\sqrt{2\epsilon_1}+2\sqrt{2(\epsilon_1+\epsilon_2)}\leq6\sqrt{2(\epsilon_1+\epsilon_2)}.\nn
}
\QED
\eprf

\subsection{Quantum Entropies}\label{app:entropies}
The Shannon entropy of a probability distribution $\{p_x\}_{x\in{\mathcal X}}$ is defined as
\begin{eqnarray}
H(\{p_x\}_{x\in{\mathcal X}}):=-\sum_{x\in{\mathcal X}}p_x\log{p_x}.\nonumber
\end{eqnarray}
The von Neumann entropy of a quantum state $\rho\in {\mathcal S}({\mathcal H})$ is defined as
\begin{eqnarray}
S(\rho):=-{\rm Tr}[\rho\log{\rho}].\nonumber
\end{eqnarray}
It is invariant under unitary transformations, i.e., $S(\rho)=S(U\rho U^\dagger)$ for any unitary $U$. With $\rho=\sum_{x\in{\mathcal X}}p_x|x\rangle\!\langle x|$ being the spectral decomposition of $\rho$, we have $S(\rho)=H(\{p_x\}_{x\in{\mathcal X}})$. The von Neumann entropy is concave, that is, we have
\alg{
S\left(\sum_jp_j\rho_j\right)\geq\sum_jp_jS(\rho_j)\nn
}
for any ensemble $\{p_j,\rho_j\}_j$. We denote the von Neumann entropy of a state $\rho$ on system $A$ interchangeably by $S(\rho^A)$ and $S(A)_\rho$. For a bipartite pure state $|\psi\rangle^{AA'}$, we have
\begin{eqnarray}
S(\psi^A)=S(\psi^{A'}).\nn
\end{eqnarray}
For a bipartite state $\rho\in{\mathcal S}({\mathcal H}^A\otimes{\mathcal H}^B)$, the von Neumann entropy satisfies the {\it subadditivity}, expressed as
\begin{eqnarray}
S(A)_\rho+S(B)_\rho\geq S(AB)_\rho.\nn
\end{eqnarray}

Define
\begin{eqnarray}
\eta(x):=\begin{cases}
x-x\log{x}&(x\leq 1/e)\\
x+\frac{1}{e}&(x\geq 1/e)
\end{cases},\nonumber
\end{eqnarray}
where $e$ is the base of the natural logarithm. For two states $\rho$ and $\sigma$ in a $d$-dimensional quantum system ($d<\infty$) such that $\|\rho-\sigma\|_1\leq\epsilon$, we have
\begin{eqnarray}
|S(\rho)-S(\sigma)|\leq\eta(\epsilon)\log{d},\nn
\end{eqnarray}
which is called the {\it Fannes inequality}\cite{fannes73}.

\subsection{Typical Sequences and Subspaces}\label{app:typsub}

Let $X$ be a discrete random variable with finite alphabet $\mathcal X$ and probability distribution $p_x={\rm Pr}\{X=x\}$ where $x\in{\mathcal X}$. A sequence ${\vec x}=(x_1,\cdots,x_n)\in{\mathcal X}^n$ is said to be {\it$\delta$-typical with respect to} $\{p_x\}_{x\in{\mathcal X}}$ if it satisfies
\begin{eqnarray}
2^{-n(H(X)+\delta)}\leq\prod_{i=1}^Np_{x_i}\leq2^{-n(H(X)-\delta)},\nn
\end{eqnarray}
where $H(X)$ is the Shannon entropy of $X$ defined by $H(X)=H(\{p_x\}_{x\in{\mathcal X}})$.
The set of all $\delta$-typical sequences is called the {\it$\delta$-typical set}, and is denoted by ${\mathcal T}_{n,\delta}$. Denoting $\prod_{i=1}^Np_{x_i}$ by $p_{\vec x}$, we have
\begin{eqnarray}
1=\sum_{{\vec x}\in{\mathcal X}^n}p_{\vec x}\geq\sum_{{\vec x}\in{\mathcal T}_{n,\delta}}p_{\vec x}\geq|{\mathcal T}_{n,\delta}|\cdot2^{-n(H(X)+\delta)},\nonumber
\end{eqnarray}
which implies
\begin{eqnarray}
|{\mathcal T}_{n,\delta}|\leq2^{n(H(X)+\delta)}.
\label{eq:cardtyp}\end{eqnarray}
The weak law of large numbers implies
\begin{eqnarray}
{\rm Pr}\{(X_1,\cdots,X_n)\in{\mathcal T}_{n,\delta}\}\geq1-\epsilon\label{eq:weaktyp}
\end{eqnarray}
for any $\epsilon,\delta>0$ and sufficiently large $n$.


Suppose the spectral decomposition of $\rho\in{\mathcal S}({\mathcal H})$ is given by $\rho=\sum_{x\in{\ca X}}p_x\proj{x}$. The {\it$\delta$-typical subspace ${\mathcal H}_{n,\delta}\subset{\mathcal H}^{\otimes n}$ with respect to $\rho$} is defined as
\begin{eqnarray}
{\mathcal H}_{n,\delta}:={\rm span}\{\ket{x_1}\cdots\ket{x_n}\in{\mathcal H}^{\otimes n}|(x_1,\cdots,x_n)\in{\mathcal T}_{n,\delta}\},\!\!\!\!\!\!\!\!\!\nonumber
\end{eqnarray}
where ${\mathcal T}_{n,\delta}$ is the $\delta$-typical set with respect to $\{p_x\}_x$. Let $\Pi_{n,\delta}$ be the projection onto ${\mathcal H}_{n,\delta}$. By definition, we have
\alg{
2^{n(S(\rho)-\delta)}{\Pi}_{n,\delta}\rho^{\otimes n}{\Pi}_{n,\delta}\leq {\Pi}_{n,\delta}.\nn
}
From (\ref{eq:cardtyp}), we have
\begin{eqnarray}
{\rm dim}\:{\mathcal H}_{n,\delta}\leq2^{n(S(\rho)+\delta)}.\nonumber
\end{eqnarray}
For any $\epsilon,\delta>0$ and sufficiently large $n$, we have
\begin{eqnarray}
&&{\rm Tr}[\Pi_{n,\delta}\rho^{\otimes n}\Pi_{n,\delta}]=\sum_{{\vec x}\in{\mathcal T}_{n,\delta}}p_{\vec x}\geq1-\epsilon\nn
\end{eqnarray}
from (\ref{eq:weaktyp}).

\section{Proof of Theorem \ref{thm:symmetrizingcost}}\label{app:proof}

We prove Theorem \ref{thm:symmetrizingcost} by separately proving the converse part $C_G(\rho)\geq D_G(\rho)$ and the direct part $C_G(\rho)\leq D_G(\rho)$. In the following, we denote ${\ca T}_G$ simply by $\ca T$, and denote the set of symmetry preserving unitaries on $A$ by ${\ca U}_{\rm SP}(G)$, and on $A^n$ by ${\ca U}_{\rm SP}(G^{\times n})$.

\subsection{Converse Part}\lsec{optimality}

The converse part of \rthm{symmetrizingcost} is formulated as
\alg{
C_G(\rho)\geq D_G(\rho).\label{eq:optimality}
}
We prove this inequality by using the following lemma.
\blmm{sgveqsg}
For any $\rho\in{\ca S}({\ca H}^A)$ and $V\in {\ca U}_{\rm SP}(G)$, we have
\alg{
S({\ca T}_G(V\rho V^\dagger))=S({\ca T}_G(\rho)).\label{eq:sgveqsg}
}
\elmm
\bprf
From Equalities (\ref{eq:defrg}) and (\ref{eq:rgeqsgmns}), we have
\alg{
&S({\ca T}_G(\rho))-S(\rho)=\min_{\sigma\in{\ca S}_{\rm sym}(G)}D(\rho\|\sigma),\nn\\
&S({\ca T}_G(V\rho V^\dagger))-S(V\rho V^\dagger)=\min_{\sigma'\in{\ca S}_{\rm sym}(G)}D(V\rho V^\dagger\|\sigma').\nn
}
The unitary invariance of the von Neumann entropy implies 
\alg{
S(\rho)=S(V\rho V^\dagger).\nn
}
We also have
\alg{
\min_{\sigma'\in{\ca S}_{\rm sym}(G)}D(V\rho V^\dagger\|\sigma')&=\min_{\sigma'\in{\ca S}_{\rm sym}(G)}D(\rho\| V^\dagger\sigma'V)\nn\\
&=\min_{\sigma\in{\ca S}_{\rm sym}(G)}D(\rho\|\sigma),\nn
}
where the first line follows from
\alg{
&D(V\rho V^\dagger\|\sigma')\nn\\
&={\rm Tr}[(V\rho V^\dagger)\log{(V\rho V^\dagger)}]-{\rm Tr}[(V\rho V^\dagger)\log{\sigma'}]\nn\\
&={\rm Tr}[\rho\log{\rho}]-{\rm Tr}[\rho\log{(V^\dagger\sigma' V)}]\nn\\
&=D(\rho\| V^\dagger\sigma'V).\nn
}
Hence we obtain (\ref{eq:sgveqsg}).\hfill$\blacksquare$
\eprf
Recalling (\ref{eq:jujiiii}), It is straightforward to generalize Lemma \ref{lmm:sgveqsg} to obtain that
\alg{
S({\ca T}_G^{\otimes n}(V\rho V^\dagger))=S({\ca T}_G^{\otimes n}(\rho))\label{eq:sgveqsgg}
}
for any $\rho\in{\ca S}(({\ca H}^A)^{\otimes n})$ and $V\in {\ca U}_{\rm SP}(G^{\times n})$.

Inequality (\ref{eq:optimality}) is proved as follows. Fix arbitrary $R>C_G(\rho)$, $\epsilon>0$ and choose sufficiently large $n$. By definition, there exist a symmetric state $\sigma_n\in{\ca S}_{\rm sym}(G^{\times n})$ and a set of unitaries $\{V_k\}_{k=1}^{2^{nR}}$ such that 
\alg{
\left\|{\ca V}_n(\rho^{\otimes n})-\sigma_n\right\|_1\leq\epsilon \label{eq:vnrhosigman}
}
for
\alg{
\ca{V}_n:\tau\rarrow\frac{1}{2^{nR}}\sum_{k=1}^{2^{nR}}V_{k}\tau V_{k}^\dagger,\nn
}
and $V_k\in{\ca U}_{\rm SP}(G^{\times n})$ for all $k\in\{1,\cdots,2^{nR}\}$. Due to the monotonicity of the trace distance and the triangle inequality, it follows from (\ref{eq:vnrhosigman}) and ${\ca T}_G^{\otimes n}(\sigma_n)=\sigma_n$ that
\alg{
&\left\|({\ca T}_G^{\otimes n}\circ{\ca V}_n)(\rho^{\otimes n})-{\ca V}_n(\rho^{\otimes n})\right\|_1\nn\\
\leq&\:\left\|({\ca T}_G^{\otimes n}\circ{\ca V}_n)(\rho^{\otimes n})-\sigma_n\right\|_1+\left\|{\ca V}_n(\rho^{\otimes n})-\sigma_n\right\|_1\nn\\
=&\:\left\|({\ca T}_G^{\otimes n}\circ{\ca V}_n)(\rho^{\otimes n})-{\ca T}_G^{\otimes n}(\sigma_n)\right\|_1+\left\|{\ca V}_n(\rho^{\otimes n})-\sigma_n\right\|_1\nn\\
\leq&\;2\left\|{\ca V}_n(\rho^{\otimes n})-\sigma_n\right\|_1\nn\\
\leq&\;2\epsilon.\label{eq:distgvandv}
}

Let $E$ be an ancillary system described by a Hilbert space $\ca{H}^E$ with dimension $2^{nR}$, and define an isometry ${\tilde V}:(\ca{H}^A)^{\otimes n}\rarrow\ca{H}^E\otimes(\ca{H}^A)^{\otimes n}$ by 
\alg{
{\tilde V}=\frac{1}{2^{nR}}\sum_{k=1}^{2^{nR}}\ket{k}^{E}\otimes V_k^{A^n}.\nn
}
Let $Z$ be a quantum system with dimension $d=\dim{{\ca H}^A}$, and let $|\psi\rangle^{AZ}$ be a purification of $\rho$, i.e., $\rho={\rm Tr}_Z[\proj{\psi}^{AZ}]$. Consider a pure state
\alg{
|{\tilde\psi}_n\rangle^{EA^nZ^n}:={\tilde V}|\psi^{\otimes n}\rangle^{A^nZ^n}.\nn
}
The von Neumann entropies for this state are calculated as follows. First, we have
\alg{
\!\!nR=\log{\dim{{\ca H}^E}}\geq S(E)_{{\tilde\psi}_n}\geq S(EZ^n)_{{\tilde\psi}_n}-S(Z^n)_{{\tilde\psi}_n},\label{eq:prfconv1}
}
where the second inequality follows from the subadditivity of the von Neumann entropy. Second, we have
\alg{
S(Z^n)_{{\tilde\psi}_n}=S(Z^n)_{\psi^{\otimes n}}=S(\rho^{\otimes n})=nS(\rho),\label{eq:prfconv2}
}
since ${\tilde\psi}_n^{Z^n}=(\psi^Z)^{\otimes n}$ and $|\psi\rangle^{AZ}$ is a purification of $\rho$. Third, we have
\alg{
S(EZ^n)_{{\tilde\psi}_n}=S(A^n)_{{\tilde\psi}_n}=S(\ca{V}_n(\rho^{\otimes n})),\label{eq:prfconv3}
}
because $|{\tilde\psi}_n\rangle$ is a pure state on $EA^nZ^n$ and
\alg{
{\tilde\psi}_n^{A^n}={\rm Tr}_{EZ^n}[|{\tilde\psi}_n\rangle\!\langle{\tilde\psi}_n|]=\ca{V}_n(\rho^{\otimes n}).\nn
}
Forth, from \rineq{distgvandv} and the Fannes inequality(\cite{fannes73}, see Appendix \ref{app:entropies}), we have
\alg{
S(\ca{V}_n(\rho^{\otimes n}))\geq S(({\ca T}_G^{\otimes n}\circ{\ca V}_n)(\rho^{\otimes n}))-n\eta(2\epsilon)\log{d},\label{eq:prfconv4}
}
where $\eta$ is a function satisfying $\lim_{x\rightarrow0}\eta(x)=0$. Finally, due to the concavity of the von Neumann entropy and Equality (\ref{eq:sgveqsgg}), we have
\alg{
S(({\ca T}_G^{\otimes n}\circ{\ca V}_n)(\rho^{\otimes n}))&\geq\frac{1}{2^{nR}}\sum_{k=1}^{2^{nR}}S({\ca T}_G^{\otimes n}(V_k\rho^{\otimes n}V_k^\dagger))\nn\\
&=S({\ca T}_G^{\otimes n}(\rho^{\otimes n}))=nS({\ca T}_G(\rho)).\label{eq:prfconv5}
}
Combining (\ref{eq:prfconv1}), (\ref{eq:prfconv2}), (\ref{eq:prfconv3}), (\ref{eq:prfconv4}) and (\ref{eq:prfconv5}), we obtain
\alg{
nR&\geq nS({\ca T}_G(\rho))-nS(\rho)-n\eta(2\epsilon)\log{d},\nn
}
which implies
\alg{
R\geq D_G(\rho)-\eta(2\epsilon)\log{d}\nn
}
due to (\ref{eq:rgeqsgmns}). Since this relation holds for any $R>C_G(\rho)$ and small $\epsilon$, we obtain \rineq{optimality}.\QED

\subsection{Direct Part}\label{sec:achieve}

The direct part of \rthm{symmetrizingcost} is formulated by the following inequality:
\alg{
C_G(\rho)\leq D_G(\rho).\label{eq:achievability}
}
We prove this inequality by showing that a rate $R$ is achievable in symmetrizing $\rho$ with respect to $G$ if $R>D_G(\rho)$. The proof proceeds along the similar line as the proof of Proposition 2 in \cite{berry05}, which uses the operator Chernoff bound as a key mathematical ingredient. In the following, we fix arbitrary $\epsilon,\delta>0$ and choose sufficiently large $n\in{\mathbb N}$.

Let ${\ca H}_{n,\delta}\subset({\ca H}^{A})^{\otimes n}$ be the $\delta$-typical subspace with respect to $\rho$, and ${\Pi}_{n,\delta}$ be the projection onto ${\ca H}_{n,\delta}$. Similarly, let ${\hat{\ca H}}_{n,\delta}\subset({\ca H}^{A})^{\otimes n}$ be the $\delta$-typical subspace with respect to ${\ca T}_G(\rho)$, and let ${\hat{\Pi}}_{n,\delta}$ be the projection onto ${\hat{\ca H}}_{n,\delta}$. As described in Appendix \ref{app:typsub}, we have
\alg{
&\Tr{}{{\Pi}_{n,\delta}\rho^{\otimes n}{\Pi}_{n,\delta}}\geq1-\epsilon,\label{eq:kettei}\\
&\Tr{}{{\hat\Pi}_{n,\delta}({\ca T}_G(\rho))^{\otimes n}{\hat\Pi}_{n,\delta}}\geq1-\epsilon\label{eq:kettei2}
}
and
\alg{
2^{n(S(\rho)-\delta)}{\Pi}_{n,\delta}\rho^{\otimes n}{\Pi}_{n,\delta}\leq {\Pi}_{n,\delta}.\label{eq:UB}
}
The dimension of ${\hat{\ca H}}_{n,\delta}$, which we denote by ${\hat D}_{n,\delta}$, is bounded from above as
\alg{
{\hat D}_{n,\delta}\leq2^{n(S({\ca T}_G(\rho))+\delta)}.\nn
}

Let us consider subnormalized states 
\alg{
&X_{n,\delta}:={\hat\Pi}_{n,\delta}{\ca T}_G^{\otimes n}({\Pi}_{n,\delta}\rho^{\otimes n}{\Pi}_{n,\delta}){\hat\Pi}_{n,\delta},\label{eq:defxnd}\\
&Y_{n,\delta}:={\tilde\Pi}_{n,\delta}X_{n,\delta}{\tilde\Pi}_{n,\delta}.\label{eq:defynd}
}
Here, ${\tilde\Pi}_{n,\delta}$ is the projection onto a subspace $\tilde{\ca{H}}_{n,\delta}\subset{\hat{\ca H}}_{n,\delta}$, spanned by the eigenvectors of $X_{n,\delta}$ with eigenvalues not smaller than $\epsilon/{\hat D}_{n,\delta}$. By definition, we have
\alg{
Y_{n,\delta}\geq\epsilon\cdot2^{-n(S({\ca T}_G(\rho))+\delta)}{\tilde\Pi}_{n,\delta},\label{eq:ynLB}
}
in addition to
\alg{
Y_{n,\delta}={\tilde\Pi}_{n,\delta}{\ca T}_G^{\otimes n}({\Pi}_{n,\delta}\rho^{\otimes n}{\Pi}_{n,\delta}){\tilde\Pi}_{n,\delta}.\label{eq:defxndd}
}

The traces of $X_{n,\delta}$ and $Y_{n,\delta}$ are bounded as follows. Due to the definition of the typical subspace, we have
\alg{
\rho^{\otimes n}={\Pi}_{n,\delta}\rho^{\otimes n}{\Pi}_{n,\delta}+(I-{\Pi}_{n,\delta})\rho^{\otimes n}(I-{\Pi}_{n,\delta}),\label{eq:thh}
}
which leads to
\alg{
({\ca T}_G(\rho))^{\otimes n}&\;={\ca T}_G^{\otimes n}(\rho^{\otimes n})\nn\\
&\;={\ca T}_G^{\otimes n}({\Pi}_{n,\delta}\rho^{\otimes n}{\Pi}_{n,\delta})\nn\\
&\quad+{\ca T}_G^{\otimes n}((I-{\Pi}_{n,\delta})\rho^{\otimes n}(I-{\Pi}_{n,\delta})).\label{eq:gtee}
}
From (\ref{eq:thh}) and (\ref{eq:kettei}), we have
\alg{
&\Tr{}{{\ca T}_G^{\otimes n}((I-{\Pi}_{n,\delta})\rho^{\otimes n}(I-{\Pi}_{n,\delta}))}\nn\\
&=\Tr{}{(I-{\Pi}_{n,\delta})\rho^{\otimes n}(I-{\Pi}_{n,\delta})}\nn\\
&=1-\Tr{}{{\Pi}_{n,\delta}\rho^{\otimes n}{\Pi}_{n,\delta}}\leq\epsilon.\label{eq:juuhachi}
}
From (\ref{eq:defxnd}), (\ref{eq:gtee}), (\ref{eq:kettei2}) and (\ref{eq:juuhachi}), we obtain that
\alg{
&\Tr{}{X_{n,\delta}}=\Tr{}{{\hat\Pi}_{n,\delta}{\ca T}_G^{\otimes n}({\Pi}_{n,\delta}\rho^{\otimes n}{\Pi}_{n,\delta}){\hat\Pi}_{n,\delta}}\nn\\
&=\Tr{}{{\hat\Pi}_{n,\delta}{\ca T}_G^{\otimes n}(\rho^{\otimes n}){\hat\Pi}_{n,\delta}}\nn\\
&\quad-\Tr{}{{\hat\Pi}_{n,\delta}{\ca T}_G^{\otimes n}((I-{\Pi}_{n,\delta})\rho^{\otimes n}(I-{\Pi}_{n,\delta})){\hat\Pi}_{n,\delta}}\nn\\
&\geq\Tr{}{{\hat\Pi}_{n,\delta}{\ca T}_G^{\otimes n}(\rho^{\otimes n}){\hat\Pi}_{n,\delta}}\nn\\
&\quad-\Tr{}{{\ca T}_G^{\otimes n}((I-{\Pi}_{n,\delta})\rho^{\otimes n}(I-{\Pi}_{n,\delta}))}\geq1-2\epsilon.\label{eq:TRXn}
}
Let us define ${\tilde\Pi}_{n,\delta}^\perp:={\hat{\Pi}}_{n,\delta}-{\tilde\Pi}_{n,\delta}$. From (\ref{eq:defxnd}) and (\ref{eq:defxndd}), we have
\alg{
X_{n,\delta}=Y_{n,\delta}+{\tilde\Pi}_{n,\delta}^\perp X_{n,\delta}{\tilde\Pi}_{n,\delta}^\perp,\nn
}
which leads to
\alg{
&\Tr{}{Y_{n,\delta}}=\Tr{}{X_{n,\delta}}-\Tr{}{{\tilde\Pi}_{n,\delta}^\perp X_{n,\delta}{\tilde\Pi}_{n,\delta}^\perp}\nn\\
&\geq1-2\epsilon-{\rm rank}[X_{n,\delta}]\cdot\frac{\epsilon}{{\hat D}_{n,\delta}}\nn\\
&\geq1-2\epsilon-{\hat D}_{n,\delta}\cdot\frac{\epsilon}{{\hat D}_{n,\delta}}=1-3\epsilon.\label{eq:TRYn}
}
Here, the second line follows from (\ref{eq:TRXn}) in addition to the definitions of ${\tilde\Pi}_{n,\delta}$ and ${\tilde\Pi}_{n,\delta}^\perp$, and the third line follows from (\ref{eq:defxnd}).

Define a linear operator
\alg{
Z_{n,\delta}({\vec g}):={\tilde\Pi}_{n,\delta}U_{\vec g}{\Pi}_{n,\delta}\rho^{\otimes n}{\Pi}_{n,\delta}U_{\vec g}^\dagger{\tilde\Pi}_{n,\delta}\label{eq:defxg}
}
for each ${\vec g}\in G^{\times n}$. From (\ref{eq:UB}), we have
\alg{
Z_{n,\delta}^*({\vec g})&:=2^{n(S(\rho)-\delta)}Z_{n,\delta}({\vec g})\label{eq:sssass}\\
&\leq{\tilde\Pi}_{n,\delta}U_{\vec g}{\Pi}_{n,\delta}U_{\vec g}^\dagger{\tilde\Pi}_{n,\delta}\nn\\
&\leq{\tilde\Pi}_{n,\delta}.\nn
}
Suppose that each $g_i$ in $\vec g$ is chosen independently according to the group invariant probability measure on $G$. Due to (\ref{eq:defxg}), (\ref{eq:juji}), (\ref{eq:jujiiii}), (\ref{eq:defxndd}) and (\ref{eq:sssass}), as an ensemble average we have
\alg{
{\mathbb E}\left[Z_{n,\delta}({\vec g})\right]={\tilde\Pi}_{n,\delta}{\ca T}_G^{\otimes n}({\Pi}_{n,\delta}\rho^{\otimes n}{\Pi}_{n,\delta}){\tilde\Pi}_{n,\delta}=Y_{n,\delta}\nn
}
and
\alg{
{\mathbb E}\left[Z_{n,\delta}^*({\vec g})\right]=2^{n(S(\rho)-\delta)}Y_{n,\delta}=:Y_{n,\delta}^*.\label{eq:defxndeltaaaa}
}
Therefore, from (\ref{eq:ynLB}) and (\ref{eq:rgeqsgmns}), the minimum nonzero eigenvalue $\lambda_{n,\delta}$ of ${\mathbb E}[Z_{n,\delta}^*({\vec g})]$ is bounded below as
\alg{
\lambda_{n,\delta}\geq\epsilon\cdot2^{-n(S({\ca T}_G(\rho))-S(\rho)+2\delta)}=2^{-n(D_G(\rho)+2\delta)}.\label{eq:lambdaLB}
}

Let $N$ be a natural number, and suppose $g_{ki}\:(1\leq k\leq N,1\leq i\leq n)$ are group elements in $G$ that are randomly and independently chosen according to the group invariant probability measure. Denote $(g_{k1},\cdots,g_{kn})$ by ${\vec g}_k$. Due to the operator Chernoff bound (see Lemma 3 in \cite{berry05}), (\ref{eq:sssass}) and (\ref{eq:defxndeltaaaa}), we have
\begin{eqnarray}
&&{\rm Pr}\left\{\frac{1}{N}\sum_{k=1}^NZ_{n,\delta}({\vec g}_k)\notin[(1-\epsilon)Y_{n,\delta},(1+\epsilon)Y_{n,\delta}]\right\}\nonumber\\
&&={\rm Pr}\left\{\frac{1}{N}\sum_{k=1}^NZ_{n,\delta}^*({\vec g}_k)\notin[(1-\epsilon)Y_{n,\delta}^*,(1+\epsilon)Y_{n,\delta}^*]\right\}\nonumber\\
&&\leq2\dim{{\tilde{\ca H}}_{n,\delta}}\cdot\exp{\left(-\frac{N\epsilon^2\lambda_{n,\delta}}{2}\right)}\nonumber\\
&&\leq2d^n\exp{\left(-\frac{N\epsilon^2\lambda_{n,\delta}}{2}\right)}\nonumber
\end{eqnarray}
for any $\epsilon\in(0,1]$, which implies that
\begin{eqnarray}
&&{\rm Pr}\left\{\left\|\frac{1}{2^{nR}}\sum_{k=1}^{2^{nR}}Z_{n,\delta}({\vec g}_k)-Y_{n,\delta}\right\|_1\leq\epsilon\left\|Y_{n,\delta}\right\|_1\right\}\nonumber\\
&&\geq1-2d^n\exp{\left(-\frac{2^{nR}\epsilon^2\lambda_{n,\delta}}{2}\right)}\label{eq:chercher}
\end{eqnarray}
for an arbitrary $R>0$. Due to (\ref{eq:lambdaLB}), if $R$ satisfies
\begin{eqnarray}
R>D_G(\rho)+3\delta,\label{eq:rdgr}
\end{eqnarray}
the R.H.S. in (\ref{eq:chercher}) is greater than $0$ for any sufficiently large $n$. Then there exists a set of group elements $\{{\vec g}_k\}_{k=1}^{2^{nR}}$ such that
\begin{eqnarray}
\left\|\frac{1}{2^{nR}}\sum_{k=1}^{2^{nR}}Z_{n,\delta}({\vec g}_k)-Y_{n,\delta}\right\|_1\leq\epsilon\left\|Y_{n,\delta}\right\|_1.\label{eq:misonolove2}
\end{eqnarray}
For each element in the set, define $V_k:=U_{{\vec g}_k}\in{\ca U}_{\rm SP}(G^{\times n})$. Construct a random unitary operation ${\mathcal V}_n$ on $A^n$ as
\alg{
{\mathcal V}_n:\tau\rightarrow\frac{1}{2^{nR}}\sum_{k=1}^{2^{nR}}V_k\tau V_k^{\dagger}.\nn
}
Substituting (\ref{eq:defxg}) and (\ref{eq:defxndd}) into (\ref{eq:misonolove2}), we obtain
\begin{eqnarray}
&\left\|{\tilde\Pi}_{n,\delta}{\mathcal V}_n({\Pi}_{n,\delta}\rho^{\otimes n}{\Pi}_{n,\delta}){\tilde\Pi}_{n,\delta}\right.\quad\quad\nn\\
&\quad\quad\quad\left.-{\tilde\Pi}_{n,\delta}{\ca T}_G^{\otimes n}({\Pi}_{n,\delta}\rho^{\otimes n}{\Pi}_{n,\delta}){\tilde\Pi}_{n,\delta}\right\|_1\leq\epsilon.\nn
\end{eqnarray}
From (\ref{eq:defxndd}), (\ref{eq:TRYn}) and Lemma \ref{lmm:ggentle} in Appendix \ref{app:TDGML}, we obtain
\begin{eqnarray}
\left\|{\mathcal V}_n({\Pi}_{n,\delta}\rho^{\otimes n}{\Pi}_{n,\delta})-{\ca T}_G^{\otimes n}({\Pi}_{n,\delta}\rho^{\otimes n}{\Pi}_{n,\delta})\right\|_1\leq12\sqrt{2\epsilon}.\nn\\
\label{eq:misonolove4}
\end{eqnarray}
Furthermore, from (\ref{eq:kettei}) and the gentle measurement lemma, we have
\alg{
\left\|\rho^{\otimes n}-{\Pi}_{n,\delta}\rho^{\otimes n}{\Pi}_{n,\delta}\right\|_1\leq2\sqrt{2\epsilon},\nn
}
which implies
\alg{
&\left\|{\mathcal V}_n(\rho^{\otimes n})-{\mathcal V}_n({\Pi}_{n,\delta}\rho^{\otimes n}{\Pi}_{n,\delta})\right\|_1\leq2\sqrt{2\epsilon},\label{eq:misonoloe1}\\
&\left\|{\ca T}_G^{\otimes n}(\rho^{\otimes n})-{\ca T}_G^{\otimes n}({\Pi}_{n,\delta}\rho^{\otimes n}{\Pi}_{n,\delta})\right\|_1\leq2\sqrt{2\epsilon}\label{eq:misonoloe2}
}
due to the monotonicity of the trace distance. From (\ref{eq:misonolove4}), (\ref{eq:misonoloe1}), (\ref{eq:misonoloe2}) and the triangle inequality, we obtain
\begin{eqnarray}
&\left\|{\mathcal V}_n(\rho^{\otimes n})-{\ca T}_G^{\otimes n}(\rho^{\otimes n})\right\|_1\leq16\sqrt{2\epsilon}.\nn
\end{eqnarray}
Noting that ${\ca T}_G^{\otimes n}(\rho^{\otimes n})\in{\ca S}_{\rm sym}(G^{\times n})$, and that $\epsilon$ in the above inequality as well as $\delta$ in (\ref{eq:rdgr}) can be arbitrarily small, we conclude that a rate $R$ is achievable if $R>D_G(\rho)$. This completes the proof of (\ref{eq:achievability}).
\QED

\section{Remark}\label{app:remark}

In the previous literatures on quantum reference frames\cite{bartlett07,gilad08,bartlett2005random} (see e.g. Section IV D of \cite{gilad08}), the authors adopted the following condition, instead of (\ref{eq:ugsug}), to define symmetry of a state $\sigma$ on a composite system $A^n$:
\alg{
U_g^{\otimes n}\sigma U_g^{\dagger\otimes n}=\sigma,\quad\forall g\in G.\label{eq:ugsugg}
}
To illustrate the difference between the two definitions, let us consider a function called the {\it regularized relative entropies of frameness} (R-REF). Corresponding to (\ref{eq:ugsug}) and (\ref{eq:ugsugg}), the R-REFs are defined respectively as
\alg{
D_G^\infty(\rho)&:=\lim_{n\rightarrow\infty}\frac{1}{n}\min_{\sigma_n\in{\ca S}_{\rm sym}(G^{\times n})}D(\rho^{\otimes n}\|\sigma_n),\nn\\
{\tilde D}_G^\infty(\rho)&:=\lim_{n\rightarrow\infty}\frac{1}{n}\min_{\sigma_n\in{\ca S}_{\rm sym}(G,n)}D(\rho^{\otimes n}\|\sigma_n),\nn
}
where we denoted the set of states satisfying (\ref{eq:ugsugg}) by ${\ca S}_{\rm sym}(G,n)$. Applying (\ref{eq:rgeqsgmns}), it is straightforward to obtain that
\alg{
D_G^\infty(\rho)&=\lim_{n\rightarrow\infty}\frac{1}{n}D(\rho^{\otimes n}\|{\ca T}_G^{\otimes n}(\rho^{\otimes n}))\nn\\
&=D(\rho\|{\ca T}_G(\rho))\nn\\
&=D_G(\rho),\nn
}
while it was proved in \cite{gilad08} that
\alg{
{\tilde D}_G^\infty(\rho)=0,\quad\forall\rho\in{\ca S}({\ca H}^A).\label{eq:goku}
} 

It has been argued in \cite{brandao2015reversible}, based on (\ref{eq:goku}), that the zeroness of the R-REF is the main obstacle in applying the framework of ``operational resource theory''\cite{horodecki2002laws,horodecki2013quantumness,brandao2015reversible} to an analysis of asymmetry. As described above, this problem is solved by adopting a definition of symmetry on composite systems in terms of (\ref{eq:ugsug}), rather than (\ref{eq:ugsugg}). Thus the framework of operational resource theory is in fact applicable to asymmetry.


\bibliography{symmetry.bib}

\end{document}